\documentclass[conference,letterpaper]{IEEEtran}

\usepackage[utf8]{inputenc} 
\usepackage[T1]{fontenc}
\usepackage{url}
\usepackage{ifthen}
\usepackage{cite}
\usepackage{amssymb}
\usepackage[cmex10]{amsmath} 
\usepackage{algorithm} 
\usepackage{algorithmic}

\newtheorem{theorem}{Theorem}[section]

\newtheorem{lemma}[theorem]{Lemma}

\begin{document}
\title{On the Maximum Number of Codewords of X-Codes of Constant Weight Three}

\author{%
  \IEEEauthorblockN{Yu Tsunoda}
  \IEEEauthorblockA{Graduate School of Science and Engineering\\
                    Chiba University\\
                    1-33 Yayoi-Cho Inage-Ku, Chiba 263-8522\\
                    Email: yu.tsunoda@chiba-u.jp}
  \and
  \IEEEauthorblockN{Yuichiro Fujiwara}
  \IEEEauthorblockA{Division of Mathematics and Informatics\\
                    Chiba University\\ 
                    1-33 Yayoi-Cho Inage-Ku, Chiba 263-8522\\
                    Email: yuichiro.fujiwara@chiba-u.jp}
}

\maketitle

\begin{abstract}
X-codes form a special class of linear maps which were originally introduced for data compression in VLSI testing and are also known to give special parity-check matrices for linear codes suitable for error-erasure channels. In the context of circuit testing, an $(m, n, d, x)$ X-code compresses $n$-bit output data $R$ from the circuit under test into $m$ bits, while allowing for detecting the existence of an up to $d$-bit-wise anomaly in $R$ even if up to $x$ bits of the original uncompressed $R$ are unknowable to the tester. Using probabilistic combinatorics, we give a nontrivial lower bound for any $d \geq 2$ on the maximum number $n$ of codewords such that an $(m, n, d, 2)$ X-code of constant weight $3$ exists.
This is the first result that shows the existence of an infinite sequence of X-codes whose compaction ratio tends to infinity for any fixed $d$ under severe weight restrictions.
We also give a deterministic polynomial-time algorithm that produces X-codes that achieve our bound.
\end{abstract}


\section{Introduction}\label{intro}
Very-large-scale integration (VLSI) testing is vital part of digital circuit production and aims to minimize the number of defective circuits due to imperfect manufacturing processes.
In typical digital circuit testing, the tester applies test patterns to the circuit under test and checks whether it outputs correct responses.
In short, VLSI testing aims to detect a discrepancy between expected and observed responses of the circuit under test.

While this type of simple comparison-based testing may seem trivial to perform, it is not always an easy job.
Indeed, the ever growing complexity of modern VLSI circuits is making the volume of required input and output patterns extremely large,
resulting in prohibitively long test time and unacceptably large tester-memory requirements \cite{McCluskey:2003}.
For this reason, various cost reduction techniques, such as scan-based logic built-in self-test (BIST) \cite{McCluskey:1986}, have been developed to make modern digital circuits testable \cite{Girard:2010}.

One important development in test cost reduction is response data compression, where a well-designed compactor hashes the expected and observed responses from the circuit under test in such a way that, if the original observed data contain unexpected bits signaling a defect, the compressed versions of expected and observed responses also exhibit discrepancies \cite{Kapur:2008}.
With this method, the amount of data we should compare becomes smaller than if we naively compare every actual output against the expected behavior bit by bit.

This idea of compressing responses is very similar to non-adaptive group testing such as pooling designs for DNA library screening \cite{Ngo:2000}. 
However, there is a key difference between group testing in bioinformatics and test compression for VLSI circuits,
which is the existence of \textit{unknowable} bits, called Xs, in the case of VLSI testing.

Ideally, the tester would like to perfectly predict the behavior of a non-defective circuit for any input pattern.
However, this is not the case in general with a complex modern circuit due to various factors such as uninitialized memory elements, bus contention, floating triple-states, and imperfect simulations \cite{Girard:2010}.
Thus, the data to be compressed during circuit testing may contain Xs, that is, logic values that are unknowable to the tester beforehand,
complicating the otherwise classic coding-theoretic problem of hashing a pair of data sets while avoiding collisions if they differ.

\textit{X-compact} \cite{Mitra:2005, Mitra:2004} is a simple method to hash responses while maintaining test quality even in the presence of Xs.
A response data compactor for X-compact is called an \textit{X-code} and restricted to a linear map \cite{Lumetta:2003}.
In the language of VLSI testing, an $(m, n, d, x)$ X-code compresses $n$-bit output from the circuit under test into $m$ bits while allowing for detecting the existence of up to $d$-bit-wise discrepancies between the observed and correct responses even if up to $x$ bits of the correct behavior are unknowable.
Hence, all else being equal, given $d$ and $x$, X-codes of higher \textit{compaction ratio} $\frac{n}{m}$ are desirable.
It is notable that X-codes are also known to be useful for error-erasure separation in coding theory \cite{Tsunoda:2018a},
where achieving larger $n$ for given $m$, $d$, and $x$ is again desirable.

Because X-codes are linear maps, we may regard an $(m, n, d, x)$ X-code as a well-designed $m\times n$ matrix $H$ over the finite field $\mathbb{F}_2$ of order $2$ that compresses $n$-dimensional vectors $\boldsymbol{a} \in \mathbb{F}_2^n$ into the corresponding $m$-dimensional vectors $H\boldsymbol{a}^T \in \mathbb{F}_2^m$.
When seen this way, desirable X-codes are those with more columns and fewer rows that achieve a higher compaction ratio.

The ability to detect discrepancies and high compaction ratio are not the only required properties, however.
For practical reasons such as power requirements, compactor delay, and wirability, it is also desirable for the number of $1$s in each column to be as small as possible \cite{Mitra:2005, Wohl:2003}.
However, a column of weight less that or equal to $x$ in an $(m, n, d, x)$ X-code makes no essential contribution to the achievable compaction ratio \cite{Lumetta:2003, Fujiwara:2010}.
Hence, our focus will be on X-codes with the largest possible number $M_{x+1}(m,d,x)$ of columns for given number $m$ of rows and other two test quality parameters $d$ and $x$ under the condition that the column weights are all restricted to $x+1$,
namely \textit{optimal} X-codes of constant weight $x+1$.

The simplest case is when $x=1$, where the compactor is only required to tolerate a single $X$.
In this case, an $(m, n, d, 1)$ X-code of constant weight $2$ can be shown to be equivalent to a graph of girth $d+2$ in graph theory \cite{Wohl:2003}.
More details on X-codes of constant weight $2$ and their connection to graph theory can be found in \cite{Wohl:2003} and references therein.

While tolerance of a single $X$ is sufficient in some cases, multiple $X$s can occur in practice.
Unfortunately, our knowledge on X-codes of constant weight $x+1$ for $x \geq 2$ is quite limited.
As we will briefly review in the next section, even for the next simplest case of $x = 2$,
the precise asymptotic behavior of $M_{3}(m,d,2)$ is only known for $d = 1$,
which is $M_{3}(m,1,2) = \Theta(m^2)$ \cite{Fujiwara:2010}.
For larger $d$, as far as the authors are aware, the only non-trivial result is an upper bound, which states that $M_{3}(m,d,2) = o(m^2)$ for $d \geq 4$, proved by using a tool from extremal graph theory \cite{Tsunoda:2018}.
While this bound suggests that the asymptotic behavior of $M_{3}(m,d,2)$ is not so simple,
whether $M_{x+1}(m,d,x)$ can be superlinear for any fixed $d$ and $x$ has remained an open problem.

Here, we make a substantial step towards understanding the asymptotic behavior of $M_{3}(m,d,2)$
by proving the existence of an infinite sequence of $(m,n,d,2)$ X-codes of constant weight $3$ whose compaction ratio tends to infinity for any $d$.
\begin{theorem}\label{mainTH1}
For any positive integer $m$,
\begin{equation*}
M_3(m, d, 2)=\begin{cases}
\Omega(m^{\frac{4}{3}}) & \text{for $d=2$},\\
\Omega(m^{\frac{5}{4}}) & \text{for $d=3$},\\
\Omega(m^{\frac{6}{5}}) & \text{for $d\ge 4$}.
\end{cases}
\end{equation*}
\end{theorem}
We first give a short and nonconstructive proof based on probabilistic combinatorics
and then provide a deterministic polynomial-time algorithm that produces X-codes that achieve the above lower bound.

In Section \ref{pre}, we briefly review the basic properties of X-codes and known results.
Section \ref{mainr} gives our new bound on $M_3(m, d, 2)$
as well as a deterministic algorithm for constructing X-codes attaining this bound that runs in time polynomial in $m$ for fixed $d$.
Section \ref{conclude} concludes this paper with some remarks.

\section{Preliminaries}\label{pre}
Here, we give a formal mathematical definition of X-codes.
Basic facts and known results are also briefly reviewed.

Let $m$, and $n$ be positive integers.
The \textit{superimposed sum} of two binary columns $\boldsymbol{a}=(a_1, a_2,\dots, a_m)^T$, $\boldsymbol{b}=(b_1, b_2, \dots, b_m)^T\in\mathbb{F}_2^{m}$ is defined to be $\boldsymbol{a}\bigvee \boldsymbol{b} = (a_1 \vee b_1, a_2 \vee b_2,\dots, a_m \vee b_m)^T$,
where $a_i\vee b_i= 0$ if $a_i=b_i=0$ and $1$ otherwise.
The \textit{addition} of $\boldsymbol{a} + \boldsymbol{b}$ between two columns $\boldsymbol{a},\boldsymbol{b}\in\mathbb{F}_2^{m}$ is assumed to be the coordinatewise sum over $\mathbb{F}_2$ as usual.
A binary column $\boldsymbol{a}$ is said to be contained in another binary column $\boldsymbol{b}$ if $\boldsymbol{a}\bigvee \boldsymbol{b}=\boldsymbol{b}$.

For positive integer $d$ and non-negative integer $x$,
an $m\times n$ binary matrix $H$ is an $(m, n, d, x)$ \textit{X-code} if the superimposed sum of any $x$ columns does not contain the \textit{addition} of any other up to $d$ columns.
The columns of an X-code are the \textit{codewords}, while the number of rows is the \textit{length}.

By definition, an $(m, n, d, x)$ X-code with $d\ge 2$ is an $(m, n, d-1, x)$ X-code.
For $d\ge 2$ and $x\ge 1$, an $(m, n, d, x)$ X-code is also an $(m, n, d+1, x-1)$ X-code \cite{Lumetta:2003}.

It is notable that the definition of an $(m, n, 1, x)$ X-code coincides with that of an \textit{$x$-disjunct matrix} \cite{Macula:1996a} of size $m\times n$ for group testing.
Disjunct matrices are also known as \textit{cover-free families} \cite{Erdos:1985} and \textit{superimposed codes} \cite{Kautz:1964}.
An $(m, n, d, 0)$ X-code forms a parity-check matrix for a linear code of length $n$, dimension at least $n-m$, and minimum distance at least $d+1$.
For $x \geq 1$, an $(m, n, d, x)$ X-code can be seen as a special parity-check matrix that can treat errors and erasures separately over an error-erasure channel \cite{Tsunoda:2018a}.

Now, to see how X-codes work in VLSI testing, let us consider the following $(4, 6, 1, 1)$ X-code $H$.
\begin{equation*}
H=\left(\begin{array}{cccccc}
1 &0 &0 &1 &1 &0 \\
1 &1 &0 &0 &0 &1 \\
0 &1 &1 &1 &0 &0 \\
0 &0 &1 &0 &1 &1 
\end{array}\right).
\end{equation*}
Assume that $\boldsymbol{a}$ and $\boldsymbol{b}$ are $6$-dimensional vectors over $\{0, 1, X\}$ that represent the observed and expected responses from the circuit under test, respectively.
Because X represents an unknowable logic value, computation involving X is defined by
$a+\text{X}=\text{X}+a=\text{X}$, $0\cdot \text{X}=\text{X}\cdot 0=0$, and $1\cdot \text{X}=\text{X}\cdot 1=\text{X}$.
VLSI testing with the X-code $H$ compares $4$-dimensional vectors $H\boldsymbol{a}^T$ and $H\boldsymbol{b}^T$ instead of $\boldsymbol{a}$ and $\boldsymbol{b}$.
The property of $H$ as a $(4,6,1,1)$ X-code guarantees that a discrepancy of no more than one bit between the actual output $\boldsymbol{a}$ and the correct output $\boldsymbol{b}$ can be detected by comparing the shrunk responses $\boldsymbol{a}$ and $\boldsymbol{b}$ even if up to $1$ bit of the correct behavior is unknowable.
For example, when $\boldsymbol{a}=(0,1,1,0,0,0)$ and $\boldsymbol{b}=(\text{X}, 1,1,1,0,0)$, the shrunk responses are $H\boldsymbol{a}^T=(0,1,0,1)^T$ and $H\boldsymbol{b}^T=(\text{X},\text{X}, 1,1)^T$.
As can be seen easily, there exists a discrepancy between the last bits of $H\boldsymbol{a}^T$ and $H\boldsymbol{b}^T$, so that we can find the circuit under test is defective.

As a linear function for compaction, all else being equal, it is desirable for an $(m, n, d, x)$ X-code to have as large $n$ as possible for given $m, d$, and $x$.
Recall that $M_{x+1}(m, d, x)$ is the maximum number $n$ for which there exists an $(m,n,d,x)$ X-code of constant weight $x+1$. 
For $d=1$ and $x=2$, it is known that 
\begin{equation*}
M_{3}(m,1,2)\le\frac{m(m-1)}{6}
\end{equation*}
with equality if and only if $m\equiv 1,3\pmod 6$ \cite{Fujiwara:2010}. 
Because $M_3(m, d, 2)\le M_{3}(m, 1, 2)$ by definition, the above upper bound holds for all $d\ge 2$ as well.
While the above inequality says that $M_{3}(m, 1, 2)=\Theta(m^2)$, it is also known that $M_{3}(m, d, 2)=o(m^2)$ for any $d\ge 4$ \cite{Tsunoda:2018}.
Therefore, we have the following theorem.
\begin{theorem}[\cite{Fujiwara:2010, Tsunoda:2018}]
It holds that
\begin{equation*}
M_3(m, d, 2)=\begin{cases}
\Theta(m^2) & \text{for $d=1$},\\
O(m^2) & \text{for $d=2,3$},\\
o(m^2) & \text{for $d\ge 4$}.
\end{cases}
\end{equation*}
\end{theorem}
Although an attempt has been made to construct matrices similar to $(m,n,d,2)$ X-codes of constant weight $3$ for large $d$ in \cite{Tsunoda:2018},
to the best of the authors' knowledge, no nontrivial lower bounds on $M_3(m, d, 2)$ are known for $d \geq 2$.

\section{Main Results}\label{mainr}
This section is divided into two subsections.
In Section \ref{submain1}, we derive a general lower bound on $M_3(m, d, 2)$ by using the probabilistic method in combinatorics \cite{Alon:2016}.
Section \ref{submain2} derandomizes the probabilistic proof to demonstrate that
an $(m, n, d, 2)$ X-code of constant weight $3$ which attains the derived lower bound can be constructed deterministically in time polynomial in $m$.

\subsection{General lower bound on $M_{3}(m, d, 2)$}\label{submain1}
Here, we prove the following general asymptotic bound.
\begin{theorem}\label{main}
For sufficiently large $m$, it holds that
\begin{equation*}
M_{3}(m, d, 2)\ge\begin{cases}
\alpha m^{\frac{4}{3}} & \text{for $d=2$},\\
\beta m^{\frac{5}{4}} & \text{for $d=3$},\\
\gamma m^{\frac{6}{5}} & \text{for $d\ge 4$},
\end{cases}
\end{equation*}
where
\begin{align*}
\alpha&=\frac{1}{4}\left(\frac{1}{1749}\right)^{\frac{1}{3}}\fallingdotseq 2.07\times 10^{-2},\\
\beta&=\frac{4}{(15)^{\frac{3}{4}}(378131)^{\frac{1}{4}}}\fallingdotseq 2.12 \times 10^{-2}, \text{ and}\\
\gamma&=\frac{5}{3}\left(\frac{5}{10606681}\right)^{\frac{1}{5}}6^{-\frac{4}{5}}\fallingdotseq 2.16\times 10^{-2}.
\end{align*}
\end{theorem}
Note that Theorem \ref{mainTH1} immediately follows from this bound.

To prove the above theorem, we will first show the following two lemmas.

\begin{lemma}\label{lemma1}
For sufficiently large $m$, there exists an $(m, cm^{\frac{4}{3}}, 2,2)$ X-code of constant weight $3$ with $c=\frac{1}{4}(\frac{1}{1749})^{\frac{1}{3}}\fallingdotseq 2.07\times 10^{-2}$.
\end{lemma}

\begin{lemma}\label{lemma2}
For any $d\ge 3$ and any $p\in[0,1]$, there exists an $(m, N(m, d, p), d, 2)$ X-code of constant weight $3$, where
\begin{align*}
N(m, d, p)=&\binom{m}{3}p-\left(\binom{m}{6}\binom{\binom{6}{3}}{3}p^3+\binom{m}{8}\binom{\binom{8}{3}}{4}p^4\right.\\
&\quad\left. +\sum_{i=3}^{d}\binom{m}{6+\lceil\frac{3i-1}{2}\rceil}\binom{\binom{6+\lceil\frac{3i-1}{2}\rceil}{3}}{i+2}p^{i+2}\right).
\end{align*}
\end{lemma}

To show these lemmas, we employ a well-known class of combinatorial designs.
A \textit{set system of order $v$} is an ordered pair $(V, \mathcal{B})$ such that $V$ is a finite set of \textit{points} with $|V|=v$ and $\mathcal{B}$ is a family of subsets of $V$, called \textit{blocks}.
The point-by-block incidence matrix of a set system $(V,\mathcal{B})$ is the binary $|V|\times |\mathcal{B}|$ matrix $H=(h_{i,j})$ such that rows and columns are indexed by points and blocks, respectively, and $h_{i,j}=1$ if the $i$th point is contained in the $j$th block and $h_{i,j}=0$ otherwise.
For a subset $\mathcal{B}'$ of $\mathcal{B}$, the \textit{odd-point union} $U$ of $\mathcal{B}'$ is defined to be
$U=\{a\in V\mid |\{B\in \mathcal{B'}\mid a\in B\}|\text{ is odd}\}$.
It is straightforward to see that an $(m, n, d, x)$ X-code whose rows and columns are indexed by $V$ and $\mathcal{B}$, respectively, is equivalent to a set system $(V,\mathcal{B})$ of order $m$ with $|\mathcal{B}|=n$ such that no union of $x$ blocks contains the odd-point union of any other $d$ or fewer blocks as a subset.
We denote by $\binom{V}{3}$ the set of $3$-subsets, called \textit{triples}, of $V$.

X-codes of constant weight $3$ can be characterized by some subsets of $\binom{V}{3}$. 
A \textit{configuration} in a set system $(V,\mathcal{B})$ is a subset $C$ of $\mathcal{B}$.
When $|C|=i$, a configuration $C$ is an $i$-configuration.
A configuration is $(d,x)$-\textit{forbidden} if it appears in no $(m, n, d, x)$ X-codes of constant weight $3$.
We denote by $\mathcal{C}_{V, d, x}$ the set of $(d,x)$-forbidden configurations on the point set $V$.
For instance, for $\{a,b,c,d\} \subseteq V$, a $3$-configuration $C=\{\{a, b, c\}, \{b, c, d\},\{a, b, d\}\}$ is $(1,2)$-forbidden, that is, $C\in\mathcal{C}_{V, 1, 2}$, because the union $\{a, b, c, d\} = \{a, b, c\} \cup \{b, c, d\}$ contains $\{a, b, d\}$ as a subset,
making it impossible to appear in an $(m,n,1,2)$ X-code on the point set $V$.

The probabilistic proof of Lemmas \ref{lemma1} and $\ref{lemma2}$ relies on the fact that it is enough to show the existence of a set system that avoids all $(d,2)$-forbidden configurations.
The set $\mathcal{C}_{V, d, 2}$ of $(d,2)$-forbidden configurations can be partitioned into the following sets of configurations:
\begin{equation*}
\mathcal{C}_{V, d, 2}=\bigcup_{3\le i\le d+2}\mathcal{C}_{V, d, 2}(i),
\end{equation*}
where $\mathcal{C}_{V, d, 2}(i)$ is the set of $i$-configurations that are $(d,2)$-forbidden.
Note that $\mathcal{C}_{V, d, 2}(1)$ and $\mathcal{C}_{V, d, 2}(2)$ are empty sets since the set system $(V, \binom{V}{3})$ consists of distinct triples.

\begin{IEEEproof}[Proof of Lemma \ref{lemma1}]
Let $V=\{1,2, \dots, m\}$.
Take a set $\mathcal{B}$ of triples by picking elements of $\binom{V}{3}$ uniformly at random with probability $p$.
Let $X=|\{C\in\mathcal{C}_{V,2,2}\mid C\subseteq \mathcal{B}\}|$ be the random variable that counts the number of $(2,2)$-forbidden configurations in $\mathcal{B}$.
Note that because discarding a triple $T$ in $\mathcal{B}$ removes the configuration containing $T$, deleting at most one triple from each $(2,2)$-forbidden configuration in $\mathcal{B}$ gives an $(m, n, 2, 2)$ X-code with $n\geq |\mathcal{B}|-X$.
Therefore, there exists an $(m, n, 2, 2)$ X-code with $n\geq\mathbb{E}(|\mathcal{B}|-X)$.

Let $\mathcal{D}_3$ be the set of $3$-configurations in $(V,\binom{V}{3})$ that consists of 6 elements of $V$. Since every configuration in $\binom{V}{3}$ consists of distinct triples, the largest number of elements of $V$ in a configuration in $\mathcal{C}_{V, 2, 2}(3)$ is $6$, that is, $\mathcal{C}_{V, 2, 2}(3)\subset \mathcal{D}_3$.
Similarly, 
$\mathcal{C}_{V, 2, 2}(4)\subset \mathcal{D}_4$, where $\mathcal{D}_4$ is the set of $4$-configurations in $(V,\binom{V}{3})$ that consists of 8 elements of $V$.
Therefore, by linearity of expectation, we have
\begin{align*}
&\mathbb{E}(|\mathcal{B}|-X)\\
&=\mathbb{E}(|\mathcal{B}|)-\mathbb{E}(X)\\
&=\binom{m}{3}p-\sum_{i=3}^{4}|\mathcal{C}_{V,2,2}(i)|p^{i}\\
&\ge\binom{m}{3}p-\sum_{i=3}^{4}|\mathcal{D}_i|p^{i}\\
&=\binom{m}{3}p-\left(\binom{m}{6}\binom{\binom{6}{3}}{3}p^3+\binom{m}{8}\binom{\binom{8}{3}}{4}p^4\right).
\end{align*}
By setting $p=2(\frac{1}{1749})^{\frac{1}{3}}m^{-\frac{5}{3}}$, the right-hand side of the above inequality is $\frac{1}{4}(\frac{1}{1749})^{\frac{1}{3}}m^{\frac{4}{3}}+f(m)$ with $f(m)=o(m^\frac{4}{3})$, as desired.
\end{IEEEproof}
It is notable that precisely counting the number of configuration in $\mathcal{C}_{V,2,2}(i)$ instead of $\mathcal{D}_i$ can only give asymptotically the same bound.

Lemma \ref{lemma2} can be obtained by essentially the same probabilistic argument as in the proof of Lemma \ref{lemma1}.
\begin{IEEEproof}[Proof of Lemma \ref{lemma2}]
Consider the random variable $Y=|\{C\in\mathcal{C}_{V, d,2}\mid C\in\mathcal{B}\}|$ that counts the number of $(d,2)$-forbidden configuration in $\mathcal{B}$ and follow the same argument as in the proof of Theorem \ref{main} to show the existence of an $(m, n, d, 2)$ X-code of constant weight $3$ with $n\geq\mathbb{E}(|\mathcal{B}|-Y)$.
Note that for any $(i+2)$-configuration $C$ in $\mathcal{C}_{V, d, 2}(i+2)$ with $3\le i\le d$, it holds that
\begin{equation*}
\left|\bigcup_{T\in C}T\right|\le 6+\left\lceil\frac{3i-1}{2}\right\rceil,
\end{equation*}
where equality holds when $i$ is even and $C$ consists of $i$ triples $A_1, A_2, B_1, B_2,\dots B_{i}$ such that $|A_1\cup A_2|=6$, $(A_1\cup A_2)\cap B_j=\phi$ for any $1\le j\le i$, and each element in the union $\bigcup_{1\le j\le i} B_j$ is contained in exactly $2$ triples of $\{B_1,B_2,\dots , B_{i}\}$, or when $i$ is odd and $C$ consists of $i$ triples $A'_1, A'_2, B'_1, B'_2,\dots B'_{i}$ such that $|A'_1\cup A'_2|=6$ and there exists exactly one triple $B$ in $\{B'_1,\dots,  B'_i\}$ such that $|(A'_1\cup A'_2)\cap B|=1$, and each element in the union $\bigcup_{1\le j\le i} B'_j$ except for an element in $(A'_1\cup A'_2)\cap B$ is contained in exactly $2$ triples of $\{B'_1,B'_2,\dots , B'_{i-2}\}$.
Therefore, by linearity of expectation, we have
\begin{align*}
&\mathbb{E}(|\mathcal{B}|-Y)\\
&=\binom{m}{3}p-\sum_{i=1}^{d}|\mathcal{C}_{V, d, 2}(i+2)|p^{i+2}\\
&\ge\binom{m}{3}p-\left(\binom{m}{6}\binom{\binom{6}{3}}{3}p^3+\binom{m}{8}\binom{\binom{8}{3}}{4}p^4\right.\\
&\phantom{SpacingSpacing}\left. +\sum_{i=3}^{d}\binom{m}{6+\lceil\frac{3i-1}{2}\rceil}\binom{\binom{6+\lceil\frac{3i-1}{2}\rceil}{3}}{i+2}p^{i+2}\right),
\end{align*}
as desired.
\end{IEEEproof}

Now, we prove Theorem \ref{main}.
\begin{IEEEproof}[Proof of Theorem \ref{main}]
By Lemma \ref{lemma1}, it holds that
$M_{3}(m, 2, 2)\ge \alpha m^{\frac{4}{3}}$,
where $\alpha=\frac{1}{4}(\frac{1}{1749})^{\frac{1}{3}}$.
For $d=3$, by setting $p=2\left(\frac{15}{378131}\right)^{\frac{1}{4}}m^{-\frac{7}{4}}$ to maximize $N(m, d, p)$, we have
$N(m, d, p)=\beta m^{\frac{5}{4}} + g(m)$ with $\beta=\frac{4}{(15)^{\frac{3}{4}}(378131)^{\frac{1}{4}}}$ and $g(m)=o(m^{\frac{5}{4}})$.
By Lemma \ref{lemma2}, it holds that
$M_3(m, 3,2)\ge \beta m^{\frac{5}{4}}$.
Similarly, for $d\ge 4$, by setting $p=2\left(\frac{30}{10606681}\right)^{\frac{1}{5}}m^{-\frac{9}{5}}$, we have
$N(m, d, p)=\gamma m^{\frac{6}{5}}+h(m)$
with $\gamma=\frac{5}{3}\left(\frac{5}{10606681}\right)^{\frac{1}{5}}6^{-\frac{4}{5}}$ and $h(m)=o(m^{\frac{6}{5}})$. 
By Lemma \ref{lemma2}, for $d\ge 4$ it holds that 
$M_3(m, d, 2)\ge \gamma m^{\frac{6}{5}}$,
as desired.
\end{IEEEproof} 

\subsection{Construction algorithm}\label{submain2}
To extract a deterministic algorithm from our probabilistic proof in the previous section, we follow the approach of \cite{Tsunoda:2018}, which uses the \textit{method of conditional expectations} \cite{Alon:2016}.

Let $T_i$, $1\le i\le \binom{m}{3}$, be the triples in $\binom{V}{3}$ in arbitrary order.
While the proof of Lemmas \ref{lemma1} and \ref{lemma2} randomly picks each triple in $\binom{V}{3}$, here we deterministically decide whether to pick $T_i$ one by one from $T_1$ through $T_{\binom{m}{3}}$.
Note that the picked triples may contain some forbidden configurations in $\mathcal{C}_{V, d, 2}$.
To remove the forbidden configurations, the final deletion process is done the same way as in the probabilistic proof by discarding at most one triple from each realized forbidden configuration.

Now we describe our derandomized algorithm in detail.
To record our decision on whether we pick a triple at each step, define
$t_i = 1$ if $T_i$ is included and $0$ otherwise.
For a given binary sequence $t_1\dots t_i$ of length $i$ and forbidden configuration $C\in\mathcal{C}_{V, d, 2}$, define $r=|\{T_j\in C\mid t_j=1, 1\le j\le i\}|$ and
\begin{equation*}
s(C)=\begin{cases}
p^{|C|-r} & \text{if $|\{T_j\in C\mid t_j=0\}|=0$},\\
0 & \text{otherwise}.
\end{cases}
\end{equation*}

Given the first $i$ decisions $t_1\dots t_i$ on the triples, assume for the moment that we pick each of remaining $\binom{m}{3}-i$ triples independently and uniformly at random with probability
\begin{equation*}
p=\begin{cases}
2(\frac{1}{1749})^{\frac{1}{3}}m^{-\frac{5}{3}} & \text{for $d=2$},\\
2\left(\frac{15}{378131}\right)^{\frac{1}{4}}m^{-\frac{7}{4}} & \text{for $d=3$},\\
2\left(\frac{30}{10606681}\right)^{\frac{1}{5}}m^{-\frac{9}{5}} & \text{for $d\ge 4$}.
\end{cases}
\end{equation*}
This is equivalent to hypothetically regarding each $t_j$ for $i+1\le j\le \binom{m}{3}$ as an independent random variable such that $t_j=1$ with probability $p$ and $0$ with probability $1-p$.

Let $\mathcal{B}=\{T_j\mid t_j=1, 1\le j \le \binom{m}{3}\}$, and define $\mathcal{X}=\{C\in\mathcal{C}_{V, d, 2}\mid C\subseteq \mathcal{B}\}$.
$\mathcal{B}$ and $\mathcal{X}$ represent the sets of picked triples and formed forbidden configurations after the hypothetical random sampling, respectively. 

If we started with fixed $i$ decisions $t_1,\dots, t_i$ and performed the random sampling for the remaining triples, the number of triples after the deletion process would be at least the conditional expectation
\begin{align*}
\mathbb{E}(|\mathcal{B}|-|\mathcal{X}|\mid t_i,\dots t_i)=&|\{T_j\mid t_j=1, 1\le j\le i\}|\\
&+\left(\binom{m}{3}-i\right)p-\sum_{C\in\mathcal{C}_{V, d, 2}}s(C).
\end{align*}
For $b=0, 1$, define
\begin{equation*}
E((t)_i, t_{i+1=b})=\mathbb{E}(|\mathcal{B}|-|\mathcal{X}|\mid t_i,\dots t_i, t_{i+1}=b),
\end{equation*}
which is the conditional expectation when the random variable $t_{i+1}$ is realized as $b$.
When $i=0$, we define
\begin{equation*}
E((t)_0, t_1=b)=\mathbb{E}(|\mathcal{B}|-|\mathcal{X}|\mid t_1=b).
\end{equation*}
Since
\begin{align*}
&\mathbb{E}(|\mathcal{B}|-|\mathcal{X}|\mid t_i,\dots t_i)\\
&\phantom{SPACEspace}=pE((t)_i, t_{i+1=1}) +(1-p)E((t)_i, t_{i+1=0}),
\end{align*}
it holds that
\begin{align*}
&\mathbb{E}(|\mathcal{B}|-|\mathcal{X}|\mid t_i,\dots t_i)\\
&\phantom{SPACESPACe}\le \max\{E((t)_i, t_{i+1=1}), E((t)_i, t_{i+1=0})\}.
\end{align*}
Therefore, by starting from no decisions on the triples and picking $T_i$ at the $i$th step if and only if $E((t)_i, t_{i+1=1})> E((t)_i, t_{i+1=0})$, we end up with at least $\mathbb{E}(|\mathcal{B}|-|\mathcal{X}|)$ triples after the deletion process, which is precisely the guaranteed number of codewords of an X-code by Theorem \ref{main}.
Algorithm \ref{algo} describes the above deterministic procedure.

\begin{algorithm}[h]
 \caption{Derandomized algorithm for Theorem \ref{main}}\label{algo}
 \begin{algorithmic}[1]
 \renewcommand{\algorithmicrequire}{\textbf{Input:}}
 \renewcommand{\algorithmicensure}{\textbf{Output:}}
 \REQUIRE Point set $V$ of cardinality $m$
 \ENSURE  $(m,|\mathcal{B}|, d, 2)$ X-code $(V, \mathcal{B})$ of constant weight $3$
  \STATE $\mathcal{B}\leftarrow \phi$
  \STATE Fix the order of $\{T_1,T_2,\dots T_{\binom{m}{3}}\} = \binom{V}{3}$ arbitrarily
  \STATE  $\mathcal{C}_{V, d, 2}\leftarrow$ set of all forbidden configurations in $\binom{V}{3}$
  \FOR{$i=1$ to $\binom{m}{3}$}
\IF{($E((t)_{i-1},t_i=1)>E((t)_{i-1},t_i=0)$)}
\STATE $\mathcal{B}\leftarrow \mathcal{B}\cup\{T_{i}\}$
\STATE $t_i \leftarrow 1$
\ELSE
\STATE $t_i \leftarrow 0$
\ENDIF
\ENDFOR
\WHILE{$\exists C\in\mathcal{C}_{V, d, 2}$ s.t. $C\subset\mathcal{B}$}
\STATE $\mathcal{B}\leftarrow \mathcal{B}\setminus \{T\}$, where $T$ is an arbitrary triple in $C$
\ENDWHILE
\RETURN $(V, \mathcal{B})$
 \end{algorithmic}
 \end{algorithm}

In the remainder of this section, we show that Algorithm \ref{algo} runs in time polynomial in $m$.
Our analysis here is quite rough but enough to show that it is efficient in a technical sense.

First, note that listing all forbidden configurations in $\mathcal{C}_{V, d, 2}$ only takes time polynomial in $m$ because $|\mathcal{C}_{V, d, 2}|=O(m^{6+\lceil \frac{3d-1}{2}\rceil})$.
The steps for picking triples require computing two conditional expectations $\binom{m}{3}$ times each.
Since computing a conditional expectation takes at most $O(|\mathcal{C}_{V, d, 2}|)$ time, the steps for picking triples can be done in $O(m^{9+\lceil \frac{3d-1}{2}\rceil})$.
Checking whether a given triple is contained in $\mathcal{B}$ takes $O(\log|\mathcal{B}|)$ time by using the binary search.
Therefore, the complexity of the final deletion process is bounded from above by $|\mathcal{C}_{V, d, 2}|\log\binom{m}{3}$.
Hence, the total run time will not exceed $O(m^{9+\lceil \frac{3d-1}{2}\rceil})$, as required.

\section{Conclusion}\label{conclude}
We have derived a lower bound on the maximum number $n$ for which an $(m, n, d, 2)$ X-code of constant weight $3$ exists.
This is the first nontrivial lower bound on $M_3(m,d,2)$ for general $d$
and demonstrates that constant-weight X-codes can substantially reduce the amount of response data
under the presence of a multi-bit discrepancy, multiple unknowable bits, and a severe constraint on fan-out
(see \cite{Fujiwara:2010} and references therein for the background on the fan-out issue).

We have also proved that such X-codes can be constructed deterministically in time polynomial in $m$.
This was done by first proving their existence through a probabilistic argument and then derandomizing it by the method of conditional expectations.
It is notable that this approach was also shown effective in \cite{Tsunoda:2018} for a more specific situation where multiple Xs are rather rare but do occur.
It would be interesting to see how widely this approach can be applied to similar problems.

Finally, it should be noted that while we have made nontrivial progress towards understanding the asymptotic behavior of $M_3(m,d,2)$, there still remains a substantial gap between the sharpest upper and lower bounds on $M_3(m,d,2)$.
In fact, for general $d$ and $x$, the problem of determining $M_{x+1}(m,d,x)$ is nearly completely open.
We hope that future work addresses these challenging areas.

\section*{Acknowledgment}
This work was supported by JSPS KAKENHI Grant Number JP18J20466 (Y.T.) and KAKENHI Grant Number JP17K12638 (Y.F.).





\end{document}